\documentclass[floatfix,floats,aps,prb,showpacs,amsfonts,amssymb]{revtex4}

\usepackage{amssymb}
\usepackage{dcolumn}
\usepackage{bm}

\usepackage{hyphenat}
\usepackage{graphicx}
\usepackage{psfrag}
\usepackage[latin1]{inputenc}
\begin{document}

\title{Bound-state third-order optical nonlinearities of embedded germanium nanocrystals}

\author{Hasan Y{\i}ld{\i}r{\i}m}
\email{hasany@fen.bilkent.edu.tr}
\affiliation{Department of Physics, Bilkent University, Ankara 06800, Turkey}
\affiliation{Nanoscience Laboratory Department of Physics, 
University of Trento Via Sommarive, 14 38100 Trento, Italy}
\author{Ceyhun Bulutay}
\email{bulutay@fen.bilkent.edu.tr}
\affiliation{Department of Physics, Bilkent University, Ankara 06800, Turkey}
\date{\today}

\begin{abstract}
Embedded germanium nanocrystals (NCs) in a silica host matrix are theoretically 
analyzed to identify their third-order bound-state nonlinearities.
A rigorous atomistic pseudopotential approach is used for determining the electronic 
structure and the  nonlinear optical susceptibilities.
This study characterizing the two-photon absorption, nonlinear refractive index, and optical switching
parameters reveals the full wavelength dependence from static up to the
ultraviolet spectrum and the size dependence up to a diameter of 3.5~nm. 
Similar to Si NCs, the intensity-dependent refractive index increases with decreasing NC diameter.
On the other hand, Ge NCs possess about an order of magnitude smaller nonlinear 
susceptibility compared to Si NCs of the same size. 
It is observed that the two-photon absorption threshold extends beyond the half band-gap value. 
This enables nonlinear refractive index tunability over a much wider wavelength range 
free from two-photon absorption.
\end{abstract}
\maketitle

\section{Introduction}
Semiconductor nanocrystals (NCs) benefit from the accumulated knowledge in semiconductor 
physics and the maturity of the semiconductor industry as well as the new opportunities 
provided by the nanoscience, hence they offer unique optical properties.\cite{gaponenko05}
In particular, Si and Ge NCs attract an increasing attention because of
their low-cost and microelectronic-compatible photonic applications ranging from light 
emitting diodes and lasers to
solar cells and other photonic devices.\cite{pavesi04, reed}
Even though Si and Ge are both Group~IV elements, there are a number of notable 
differences between them such as the band edge effective mass of the carriers are smaller for Ge, 
whereas the dielectric constant is larger which results in bulk exciton radius 
about five times larger for Ge compared to Si.\cite{yu-cardona}
As a result, the confinement effects will be felt starting from larger sizes. 
Moreover, Ge is a weakly indirect band-gap semiconductor with the direct to indirect band-gap 
ratio being 1.2, in contrast to 2.9 in Si. 
Furthermore, the narrower band-gap of the bulk as well as the NC Ge can be preferred in certain 
applications to harvest the near infrared part of the spectrum.\cite{healy92}
Finally, the proximity of direct-gap optical transitions in bulk Ge to the fiber optic communication 
wavelength of 1.5~$\mu$m range is particularly important. This has recently stimulated extensive 
interest; notably tensile-strained Ge photodetectors on Si platform has been demonstrated~\cite{liu05} and a 
tensile-strained Ge-based laser is proposed.\cite{liu07} If the latter can also be experimentally demonstrated, 
this will mark the dawn of the germanium photonics era.

For all-optical switching and sensor protection applications\cite{boyd03} as well as in the 
absorption of the subband-gap light for the possible solar cell applications,\cite{thurpke}
the nonlinear refractive index coefficient also known as the optical Kerr index, $n_{2}$, and 
two-photon absorption coefficient, $\beta$, are the two crucial third-order optical nonlinearities 
which play an important role.
Recent experiments show that Ge NCs have enhanced  third-order
optical nonlinearities.\cite{dowd99,jie00,li01,wan03,razzari06,gerung06}
However, differences in the sample preparation methods, the choice of the matrix, the excitation 
laser wavelength and the size distribution of the NCs contribute to the wide 
variance within these results as shown in Table~I.
A number of these investigations have observed two characteristic 
temporal nonlinear response contributions, distinguished as fast and slow, but there is no  
quantitative agreement among themselves.\cite{dowd99,li01,razzari06}
Regarding the origin of the nonlinear response, some of these reports have stressed the role of the 
excited-state contribution produced by the linear absorption,\cite{jie00,li01,razzari06} also the 
involvement of the trap/defect states was addressed.\cite{dowd99,razzari06}
Undoubtedly, more experiments are needed to reach to a coherent understanding.
On the other hand, to the best of our knowledge, there is no theoretical study identifying 
the wavelength and size dependences of $n_{2}$ and
$\beta$ in Ge NCs.  Therefore, a rigorous theoretical work may guide and inspire further experimental 
studies on the foregoing investigations. Moreover, it would help in assessing the potential 
role of Ge NCs, if any, in nonlinear device applications mentioned above.

In this paper, our aim is to present such a theoretical account concerning $n_{2}$ and
$\beta$ in Ge NCs revealing their size scaling and wavelength dependence from static 
up to ultraviolet region together with a comparison with Si NCs. 
Furthermore, we deal with NCs embedded in a wide band-gap matrix representing silica 
which is the most common choice in the actual structures as can be observed in Table~I.
Since we do not consider any interface defects, strain and thermal effects or the compounding contribution of the 
excited carriers through linear absorption to the nonlinear processes, our results may serve as a benchmark 
of the ideal Ge NC bound-state ultrafast third-order nonlinearities.
In Sec.~II we describe the theoretical approach for the electronic structure and the expressions 
for nonlinear optical quantities. The results and discussions are provided in Sec.~III followed by
a brief conclusion.

\begin{table}[htb]
\caption{The summary of existing experimental studies on the third-order nonlinear optical parameters, 
$n_2$ and $\beta$ of Ge NCs.
The sample diameter, $D$, laser excitation wavelength, $\lambda_{exc}$, host matrix and sample preparation 
information are provided. Unspecified data is left as blank.}
\begin{tabular}{c c c c c c c}
\hline
\hline
Reference &  $D$~(nm)   & $\lambda_{exc}$~(nm)	&  $n_2$~(cm$^2$/GW) &	$\beta$~(cm/GW) & Matrix & Preparation    \\
\hline
 Ref.\onlinecite{dowd99}   & 3 & 800 &  2.7-6.9$\times 10^{-7}$& & silica & ion implantation \\
 Ref.\onlinecite{jie00}        & 6$\pm$1.8 & 532 &2.6-8.2$\times 10^{-3}$ &190-760 & silica & cosputtering \\
 Ref.\onlinecite{li01}          & 6$\pm$1.8 & 780 &1.5-8$\times 10^{-6}$ & 0.18-0.68 & silica & cosputtering \\
 Ref.\onlinecite{wan03}     & 5.8, 6.4 & 532 & & 95.4, 143& alumina & E-beam coevaporation \\
 Ref.\onlinecite{razzari06} &  & 800 &1$\times 10^{-6}$& 0.04  & silica & PECVD \\
 Ref.\onlinecite{gerung06} & 5$\pm$2 & 820 & &1190-1940 & solution & chemical synthesis \\
\hline
\hline
\end{tabular}
\end{table}

\section{Theory}
\begin{figure}
\includegraphics[width=9 cm]{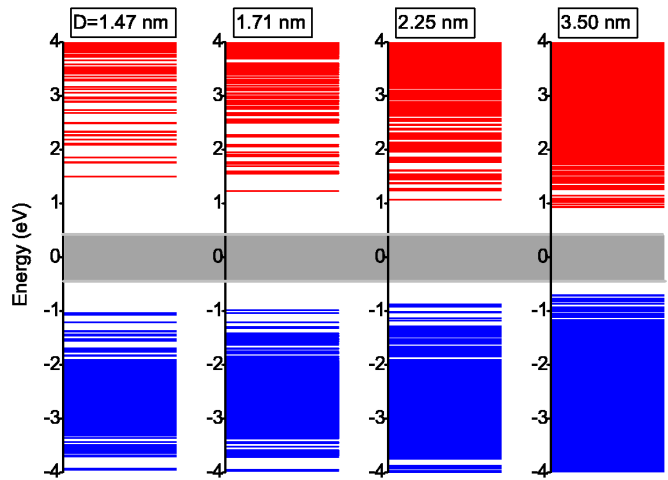}
\caption{\label{fig1} The energy levels of Ge NCs for different NC sizes. 
All plots use the same energy reference where the bulk Ge band gap is marked by the gray band.}
\end{figure}

The electronic structure of nanoclusters are accurately obtained routinely by means of density functional 
theory-based \textit{ab initio} techniques.\cite{martin-book}
However, a several nanometer-diameter NC system including the embedding host matrix contains
thousands of atoms. This large number currently precludes the use of such \textit{ab initio} 
pseudopotential plane wave techniques. An alternative route is based on the use of semi-empirical 
pseudopotential description of the atomic environment\cite{chelikowsky}
in conjunction with the linear combination of Bloch bands as the expansion basis.\cite{ninno85,wang97}
In the case of embedded Si and Ge NCs, this 
yields results in well agreement with experimental data for the interband, intraband optical 
absorption\cite{bulutay07} and the Auger recombination and carrier multiplication.\cite{sevik08}
All of these are governed by quantum processes taking place over several electronvolt energy 
range. As a matter of fact, this feature forms an essential support for applying the approach to the 
characterization of the third-order 
nonlinear susceptibilities up to a photon energy of 4~eV. We refer to our previous work for 
further details on the electronic structure.\cite{bulutay07} The corresponding electronic 
structure for embedded Ge NCs of different sizes are shown in Fig.~\ref{fig1}. The evolution of the 
effective gap, $E_G$ towards the bulk value (as marked by the gray band) can be observed as the 
NC diameter increases which is the well-known quantum size effect.

In this work, the electromagnetic interaction Hamiltonian is taken as
$-e\textbf{r}\cdot\textbf{ E}$, in other words, the length gauge is used. 
The third-order optical nonlinearity expressions based on
the length gauge have proved to be successful in atomic-like
systems\cite{boyd03} but not in bulk systems because the position operator introduces 
certain difficulties which can actually be overcome.\cite{aversa95}
Nevertheless, for bulk systems the velocity gauge has been preferred which on the other hand 
possesses unphysical divergent terms at zero frequency (not present in
the length gauge) that poses severe obstacles in evaluating the nonlinear
optical expressions.\cite{sipe00} Hence, we have preferred the length gauge for the evaluation of
the third-order optical expressions due to resemblance of the band structure of 
NCs to atomic-like systems (cf. Fig.~\ref{fig1}).
The susceptibility expression is obtained through perturbation solution of the
density matrix equation of motion.\cite{boyd03} Throughout this work, we distinguish 
the quantities which refer to unity volume filling factor by an overbar, where
$f_{v}=V_{\mbox{\begin{scriptsize}NC\end{scriptsize}}}/V_{\mbox{\begin{scriptsize}SC\end{scriptsize}}}$ 
is the volume filling factor of the NC in the matrix,
$V_{\mbox{\begin{scriptsize}NC\end{scriptsize}}}$ and $V_{\mbox{\begin{scriptsize}SC\end{scriptsize}}}$ 
are the volumes of the NC and supercell,
respectively. The final expression is given by\cite{boyd03}
$$
\overline{\chi}^{(3)}_{dcba}(-\omega_{3};\omega_{\gamma},\omega_{\beta},
\omega_{\alpha})\equiv \frac{\chi^{(3)}_{dcba}(-\omega_{3};\omega_{\gamma},
\omega_{\beta},\omega_{\alpha})}{f_{v}},
$$
\begin{equation}\label{chi3}
=\frac{e^{4}}{V_{\mbox{\begin{scriptsize}NC\end{scriptsize}}}
\hbar^{3}}\textbf{S}\sum_{lmnp}\frac{r^{d}_{mn}}{\omega_{nm}-\omega_{3}}
\left[\frac{r^{c}_{nl}}{\omega_{lm}-\omega_{2}}\left(\frac{r^{b}_{lp}r^{a}_{pm}f_{mp}}
{\omega_{pm}-\omega_{1}}-\frac{r^{a}_{lp}r^{b}_{pm}f_{pl}}{\omega_{lp}-\omega_{1}}\right)-
\frac{r^{c}_{pm}}{\omega_{np}-\omega_{2}}  \left(\frac{r^{b}_{nl}r^{a}_{lp}f_{pl}}
{\omega_{lp}-\omega_{1}}-\frac{r^{a}_{nl}r^{b}_{lp}f_{ln}}
{\omega_{nl}-\omega_{1}}\right)\right],
\end{equation}
where the subscripts $\{a,b,c,d\}$ refer to Cartesian indices,
$\omega_{3}\equiv\omega_{\gamma}+\omega_{\beta}+\omega_{\alpha}$,
$\omega_{2}\equiv\omega_{\beta}+\omega_{\alpha}$,
$\omega_{1}\equiv\omega_{\alpha}$ are the input frequencies, 
$\textbf{r}_{nm}$ is the matrix element of the position
operator between the states $n$ and $m$, $\hbar\omega_{nm}$ is the
difference between energies of these states, \textbf{S} is the
symmetrization operator,\cite{boyd03} indicating that the following
expression should be averaged over the all possible permutations of
the pairs $(c,\omega_{\gamma})$, $(b,\omega_{\beta})$, and
$(a,\omega_{\alpha})$, and finally $f_{nm}\equiv f_{n}-f_{m}$ where 
$f_{n}$ is the occupancy of the state $n$.  The
$\textbf{r}_{nm}$ is calculated for $m\neq
n$ through $\textbf{r}_{nm}=\frac{\textbf{p}_{nm}}{im_{0}\omega_{nm}}$ where
$m_{0}$ is the free electron mass, and $\textbf{p}_{nm}$ 
is the momentum matrix element. Hence, after the solution of the electronic structure, 
the computational machinery is based on the matrix elements of the standard momentum 
operator, $\textbf{P}$, the calculation of which trivially reduces to simple summations.

The above susceptibility expression is evaluated without any approximation taking into 
account all transitions within the 7~eV range. This enables a converged spectrum 
up to the ultraviolet spectrum. In the case of relatively large NCs the number of states falling 
in this range becomes excessive making the computation quite demanding.
For instance, for the 3~nm NC the number of valence and conduction states 
(without the spin degeneracy) become 3054 and 3314, respectively.
As another technical detail, the perfect $C_{3v}$ symmetry of the spherical 
NCs \cite{bulutay07} results in an energy spectrum with a large number of degenerate states. However,
this causes numerical problems in the computation of the susceptibility expression 
given in Eq.~(\ref{chi3}). This high symmetry problem can be practically removed by introducing two widely 
separated vacancy sites deep inside the matrix. Their sole effect is to 
introduce a splitting of the degenerate states by less than 1~meV.

When solids are excited with light having a  frequency
below the band-gap at intensities high enough, third-order changes  in the refractive index and
the absorption are observed due to the virtual excitations of the
bound charges. Accounting for these effects, the refractive index
and the absorption become, respectively,
$n=n_{0}+n_{2}I$, $\alpha=\alpha_{0}+\beta I$ ,
where $n_{0}$ is the linear refractive index, $\alpha_{0}$ is the linear absorption 
coefficient, and \textit{I} is the intensity of the light.
$\overline{n}_{2}$ is proportional to $\textrm{Re}\left\{ \overline{\chi}^{(3)} \right\}$, and is
given by\cite{bahae90}
\begin{equation}\label{n2}
\overline{n}_{2}(\omega)=\frac{\textrm{Re}\left\{ \overline{\chi}^{(3)}(-\omega;\omega,
-\omega,\omega)\right\}}{2n^{2}_{0}\epsilon_{0}c}\, ,
\end{equation}
where \textit{c} is the speed of light. Similarly, $\overline{\beta}$ is given
by\cite{bahae90}
\begin{equation}\label{beta}
\overline{\beta}(\omega)=\frac{\omega\textrm{Im}\left\{ \overline{\chi}^{(3)}(-\omega;\omega,-\omega,
\omega)\right\}}{n^{2}_{0}\epsilon_{0}c^{2}}\, ,
\end{equation}
where $\omega$ is the angular frequency of the light. Note that
Eqs.~(\ref{n2}) and (\ref{beta}) are valid only in the case of
negligible absorption. The degenerate two-photon absorption cross
section $\overline{\sigma}^{(2)}(\omega)$ is given by\cite{boyd03}
\begin{equation}
\overline{\sigma}^{(2)}(\omega)\equiv\frac{\sigma^{(2)}(\omega)}{f_{v}}=\frac{8\hbar^{2}\pi^{3}
e^{4}}{n^{2}_{0}c^{2}}\sum_{i,f}\left|\sum_{m}\frac{\textbf{r}_{fm}\textbf{r}_{mi}}
{\hbar\omega_{mi}-\hbar\omega-i\hbar\Gamma}\right|^{2}
\delta(\hbar\omega_{fi}-2\hbar\omega),
\end{equation}
where $\Gamma$ is the inverse of the lifetime; the corresponding full width
energy broadening  of 100~meV is used throughout this work. 
The sum over the intermediate states, $m$, requires all interband and intraband
transitions. As we have mentioned previously we compute such expressions without any approximation by including all 
states that contribute to the chosen energy window. Finally,
$\overline{\sigma}^{(2)}(\omega)$ and $\overline{\beta}$ are
related to each other through
$\overline{\beta}=2\hbar\omega\overline{\sigma}^{(2)}(\omega)$.

Another important factor is the so-called local field effect (LFE) which arise in composite 
materials of different optical properties; the LFEs lead to a correction factor 
in the third-order nonlinear optical expressions given by,\cite{sipe02}
$
L=\left(\frac{3\epsilon_{h}}{\epsilon_{\mbox{\begin{scriptsize}NC\end{scriptsize}}}+2\epsilon_{h}}\right)^{2}
\left|\frac{3\epsilon_{h}}{\epsilon_{\mbox{\begin{scriptsize}NC\end{scriptsize}}}+2\epsilon_{h}}\right|^{2}
$
where $\epsilon_{h}$ and $\epsilon_{\mbox{\begin{scriptsize}NC\end{scriptsize}}}$ are  the dielectric
functions of the host matrix and the NC, respectively.  
We fix the local field correction at its \textit{static} value, since 
when the correction factor is a function of the  wavelength 
it brings about unphysical negative absorption regions at high energies. 
Further discussions of our model are available in our previous work.\cite{yildirim08}

\begin{figure}
\includegraphics[width=11 cm]{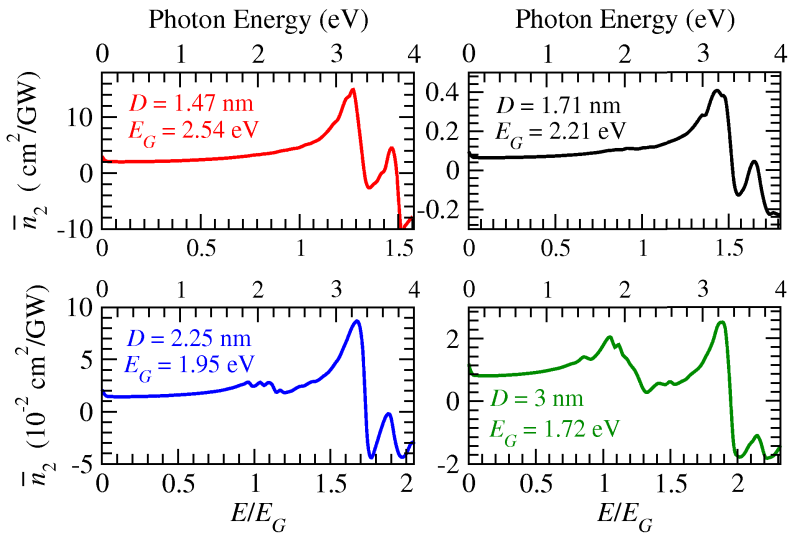}
\caption{\label{fig2}Optical Kerr index at unity filling factor, 
$\overline{n}_{2}$ in Ge NCs as a function of the photon energy for different NC sizes.
The vertical labels in the ordinates apply to both plots in the same row and the horizontal labels in the lower 
and upper abscissas apply to both plots in the same column.}
\end{figure}

\section{Results and Discussion}
We have performed extensive computations on Ge NCs with six different diameters, 
namely, $D=1.13$, 1.47, 1.71, 2.25, 3, and 3.5~nm. 
For $D=1.13$, and 3.5~nm sizes which correspond to smallest and largest diameters, 
we have calculated the nonlinearities at certain important laser wavelengths. As for the rest, 
the nonlinearities are computed at all frequencies up to 4~eV.  
For generality, we quote the unity-filling-factor values denoted by an overbar which can 
trivially be converted to any specific realization, however, we should caution that the actual 
amount of Ge atoms forming the NCs is usually a small fraction of the overall excess Ge atoms 
most of which disperse in the matrix without aggregating into a significant NC. 
In accounting for the LFE the host matrix 
is assumed to be silica which is the most common choice (cf., Table~I).

\begin{figure}
\includegraphics[width=12 cm]{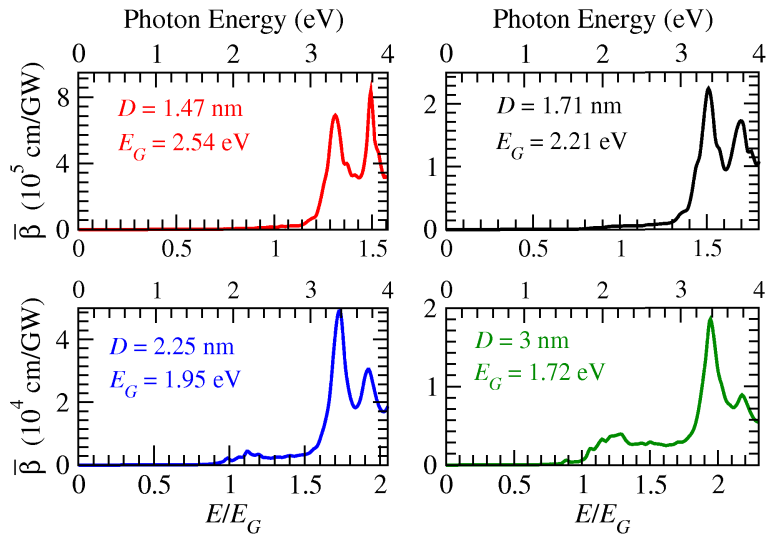}
\caption{\label{fig3} Two-photon absorption coefficient at unity filling factor, $\overline{\beta}$ 
in Ge NCs as a function of the photon energy for different NC sizes.
The vertical labels in the ordinates apply to both plots in the same row and the horizontal labels in the lower 
and upper abscissas apply to both plots in the same column.}
\end{figure}
\begin{figure}
\includegraphics[width=15 cm]{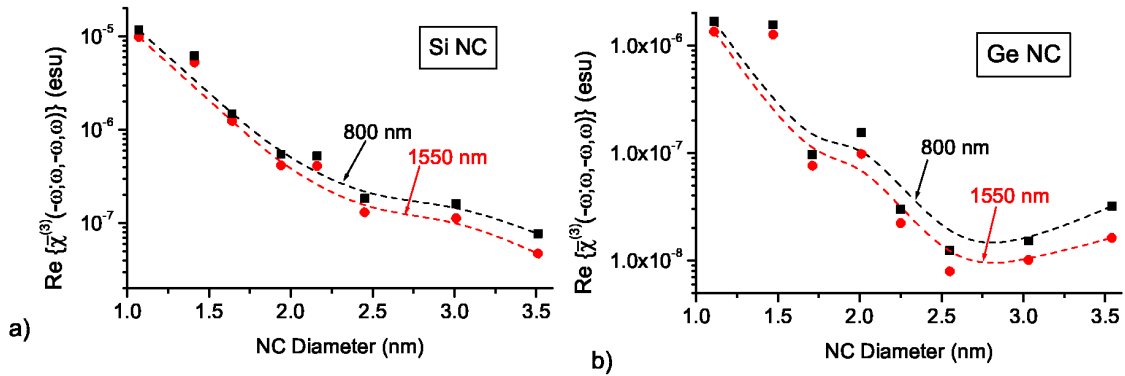}
\caption{\label{fig4} The size-scaling of the real part of the third-order susceptibility evaluated at two 
different wavelengths, 1550~nm and 800~nm, for (a) Si and (b) Ge NCs at unity filling factors. The NCs are
embedded in silica matrix. The dashed lines are guide to the eye for indicating the 
overall scaling trend.}
\end{figure}

Under these conditions, $\overline{n}_{2}$ is plotted in Fig.~\ref{fig2} as a function 
photon energy both in eV (upper abscissas) and in units of effective gap, $E_G$ (lower abscissas). 
As expected, below half $E_G$ (i.e., in the transparency region), there is a monotonous behavior and above 
this value the resonances take over. 
An intriguing observation is about the sign of ${n}_{2}$ . In general, bulk semiconductors change 
the sign of ${n}_{2}$ above their half $E_G$ values.\cite{bahae91} However, in Fig.~\ref{fig2}
we observe that in Ge NCs this takes place at even beyond $E_G$; 
for the $D=2.25$ and  3~nm NCs,  the emergence of a new resonance is seen to develop which can 
possibly reproduce this sign change at lower than $E_G$ values for the larger NCs. 
This negative sign of ${n}_{2}$ is known to be caused by both the two-photon absorption and 
the ac Stark effect.\cite{bahae91} In the size range considered in this work, we believe 
the former to be more dominant in the negative sign of ${n}_{2}$.
For $D=3$ nm NC, and below 1~eV, ${n}_{2}$ is of the order of $f_{v}\times 10^{-2}$~cm$^{2}$/GW. 
This value is  much larger than the bulk value.\cite{bahae91}
When compared to the available Ge NC measurements in Table~I, this is in very good agreement 
with Jie \emph{et al.}\cite{jie00} given the discrepancy in the NC size and the host matrix. On the other hand, 
other measurements are about three orders of magnitude lower for the same quantity.

\begin{figure}
\includegraphics[width=11 cm]{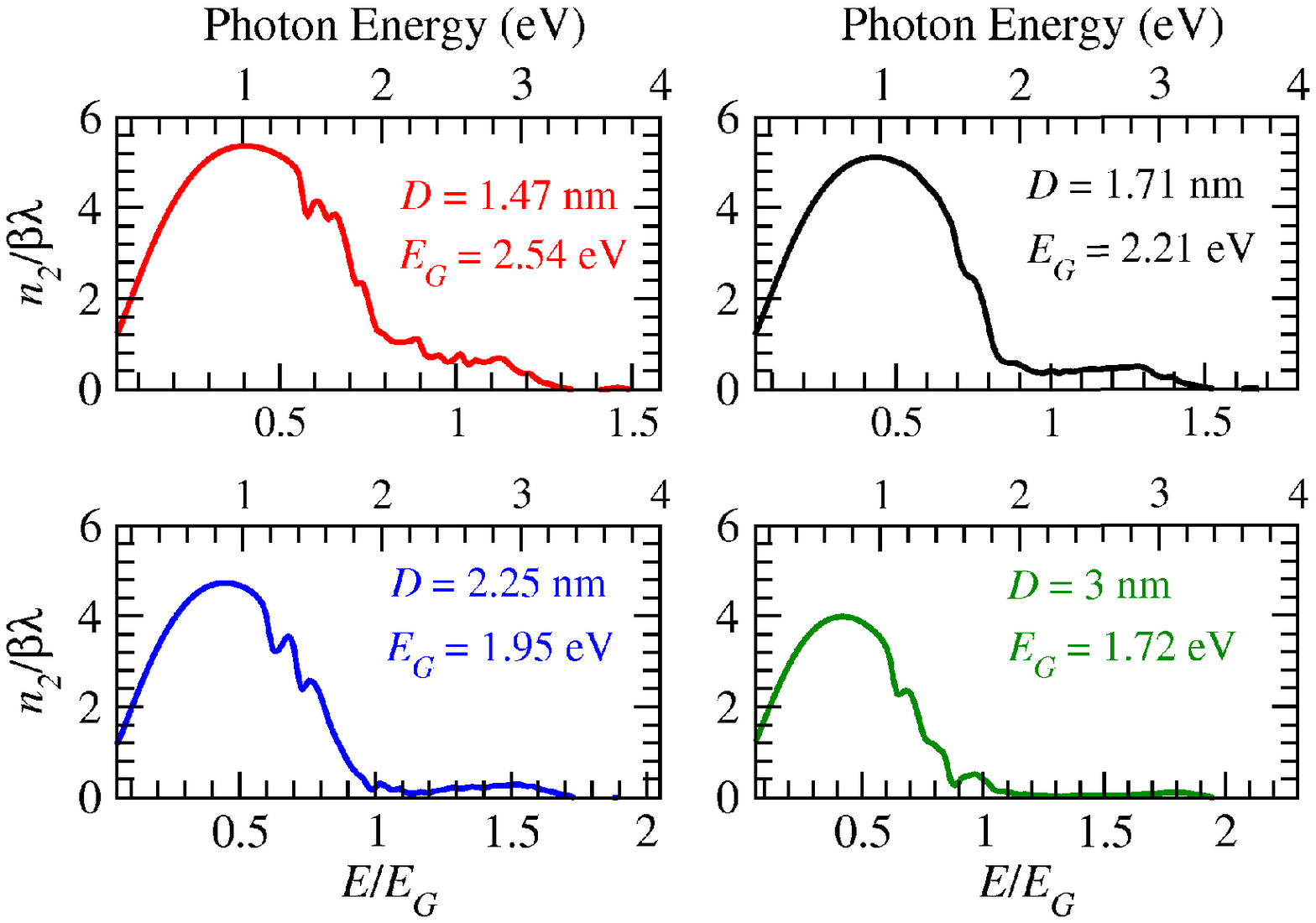}
\caption{\label{fig5} Optical switching parameter, $n_{2}/\beta\lambda$ in Ge NCs as a function of the 
photon energy for different NC sizes. The lower and upper abscissas apply to both plots in the same column.}
\end{figure}

The situation is similar in the case of $\overline{\beta}$ which is plotted against the photon energy 
in Fig.~\ref{fig3}. An important observation is that the two-photon absorption onset lies further beyond the 
half band gap value which is possibly a manifestation of the indirect band gap nature of the 
core medium.
We should note that $\overline{\beta}$ is nonzero (albeit very small) down to
static values due to band tailing\cite{lloyd} which is represented as in our previous work on 
Si NCs\cite{yildirim08} through the Lorentzian energy broadening parameter, $\hbar\Gamma$,
of 100~meV at full width.
For $D=3$ nm NC, $\beta$ has a value $4 f_{v}\times 10^{3} $~cm/GW 
around 2~eV. 
This value is very high compared to the corresponding bulk value.\cite{rauscher97}
When typical volume filling fraction is taken into account, our values are again in order of magnitude 
agreement with Gerung \emph{et al.}\cite{gerung06} and Jie \emph{et al.}\cite{jie00} both of which are for 
somewhat larger NCs. 
It should be noted that there is an outstanding disagreement among the experimental data; 
for instance, the two most recent experimental data measured at very close photon 
energies differ by five orders of magnitude.\cite{razzari06,gerung06} 
This emerging picture about the large discrepancy on the $n_2$ and $\beta$ values 
calls for further experimental investigations, especially probing the ultrafast response.

The comparison of the size-scaling trends of the real part of the third-order susceptibility for Si and Ge NCs 
are shown in Fig.~\ref{fig4}. Two different wavelengths are used, 1550~nm and 
800~nm both of which fall below the band gap, hence they do not experience any linear absorption. 
Again unity volume filling factor values are quoted.
It can be observed that for both Si and Ge NCs, there is a common enhancement trend (dashed lines) 
as the size is reduced especially below 2.5~nm.
The oscillations for certain diameters is common to both materials, however, they are more pronounced in the Ge NCs.
This corroborates with the size scaling of the Auger and carrier multiplication lifetimes.\cite{sevik08}
Another important finding is that third-order susceptibility of Si NCs are more than 20 times 
larger compared to Ge NCs of the same size embedded in the same host matrix.

For optical switching and modulation applications, one needs large tunability of the 
refractive index, such as through the optical Kerr effect without an appreciable 
change in the attenuation. Hence, as the figure of merit, $n_{2}/\beta\lambda$  is 
proposed.\cite{bahae91} In our previous work, we have 
observed that the Si NCs possess much superior figure of merit parameters 
compared to bulk Si.\cite{yildirim08} 
Fig.~\ref{fig5} shows that Ge NCs closely resemble the results of Si NCs.\cite{yildirim08} 
Both Si and Ge NCs benefit from the significant blue shift of the onset of the two-photon absorption from the half 
$E_{G}$ value, which enables the design of such switching or modulation elements over an extended 
wavelength range.
A further indirect advantage of this could be the suppression of the optical loss introduced by two-photon 
absorption generated carriers at moderately high pump powers which was a major concern in 
silicon Raman amplifiers.\cite{liang}

\section{Conclusions}
In summary, we have investigated the wavelength and size dependence
of the third-order optical nonlinearities in Ge NCs, where
our results can serve as a benchmark of the bound-state contribution reflecting
the ultrafast response of an unstrained perfect sample with no size dispersion.
Our computed values for $n_2$ and $\beta$  are in agreement with some of 
the existing experimental data which contain several orders 
of magnitude disagreement among themselves.
We observe that below the band gap, there is a common enhancement trend of both 
the real and imaginary parts of the third-order susceptibility as the NC size is reduced.
Another important finding is that third-order susceptibility of 
Ge NCs are about an order of magnitude smaller 
compared to Si NCs of the same size and embedded in the same 
dielectric environment. As in the case of Si NCs, the two-photon absorption threshold extends 
beyond the half band-gap value. This enables nonlinear refractive index tunability 
over a much wider wavelength range free from two-photon absorption.
As a final remark, our investigation calls for further experimental work especially to probe  
the ultrafast third-order nonlinear response of Ge NCs.

\begin{acknowledgments}
This work has been supported by the Turkish Scientific and Technical 
Council T\"{U}B\.ITAK with the project
number 106T048 and by the European FP6 Project SEMINANO
with the contract number NMP4 CT2004 505285. The authors would like to 
thank Dr. Can U\u{g}ur Ayfer 
for the access to Bilkent University Computer Center facilities. 
H.Y. acknowledges T\"{U}B\.ITAK-B\.IDEB for the
financial support.
\end{acknowledgments}

\newpage

\newpage

\end{document}